\documentclass[10pt,conference]{IEEEtran}
\IEEEoverridecommandlockouts

\usepackage{cite}
\usepackage{algpseudocode}
\usepackage{graphicx}
\usepackage{amsmath}
\usepackage{amsfonts}
\usepackage{algorithm}
\usepackage{float}

\newtheorem{theorem}{Theorem}

\newtheorem{lemma}{Lemma}

\newcommand{\T}{\textnormal{T}}
\newcommand{\R}{\textnormal{R}}
\let\ss= \scriptscriptstyle
\begin{document}
	\title{Analysis of Molecule Harvesting by Heterogeneous Receptors on MC Transmitters}
	\author{\IEEEauthorblockN{Xinyu Huang\IEEEauthorrefmark{1}, Yu Huang\IEEEauthorrefmark{2}, Miaowen Wen\IEEEauthorrefmark{3}, Nan Yang\IEEEauthorrefmark{1}, and Robert Schober\IEEEauthorrefmark{4}
		}
		\IEEEauthorblockA{\IEEEauthorrefmark{1}School of Engineering, Australian National University, Canberra, ACT 2600, Australia}
		\IEEEauthorblockA{\IEEEauthorrefmark{2}School of Electronics and Communication Engineering, Guangzhou University, Guangzhou, China}
		\IEEEauthorblockA{\IEEEauthorrefmark{3}School of Electronic and Information Engineering, South China University of Technology, Guangzhou, China}
	\IEEEauthorblockA{\IEEEauthorrefmark{4}Friedrich-Alexander University Erlangen-N\"{u}rnberg, 91058 Erlangen, Germany}}
	\maketitle
	\begin{abstract}
		This paper designs a molecule harvesting transmitter (TX) model, where the surface of a spherical TX is covered by heterogeneous receptors with different sizes and arbitrary locations. If molecules hit any receptor, they are absorbed by the TX immediately. Within the TX, molecules are stored in vesicles that are continuously generated and released by the TX via the membrane fusion process. Considering a transparent receiver (RX) and molecular degradation during the propagation from the TX to the RX, we derive the molecule release rate and the fraction of molecules absorbed by the TX as well as the received signal at the RX. Notably, this analytical result is applicable for different numbers, sizes, and locations of receptors, and its accuracy is verified via particle-based simulations. Numerical results show that different vesicle generation rates result in the same number of molecules absorbed by the TX, but different peak received signals at the RX.
	\end{abstract}
	
	\section{Introduction}
	Ubiquitous connectivity, encompassing nanoscale networks, has been recognized as a cutting-edge and underpinning usage scenario for the sixth-generation (6G) and beyond systems \cite{zhang20196g}. Notably, molecular communication (MC) stands out as an efficient technique for nanoscale communications, utilizing molecules carrying chemical signals to exchange information. The diffusion mechanism in MC is favored for signal transmission as no external energy is required \cite{farsad2016comprehensive}. However, molecule generation and release from the transmitter (TX) is an energy consuming process in general. For example, in biology, cells produce signaling molecules by consuming adenosine triphosphate (ATP) \cite{alberts2015essential}. As nanomachines usually operate in resource-constrained environments, such as within living organisms, minimal energy consumption is crucial for prolonging their functionalities. Moreover, energy-efficient nanomachines can perform tasks effectively, as they are able to manage power consumption and distribute resources, resulting in improved overall performance. Motivated by this, some previous studies proposed different methods to improve the energy efficiency of MC systems. In \cite{qiu2017bacterial}, the authors used bacteria as mobile relays and analyzed their information delivery energy efficiency. In \cite{deng2016enabling,guo2017smiet}, the authors proposed a simultaneous molecular information and energy transfer technique to reduce the cost of molecule synthesis, where a relay can decode the received information as well as generate molecules for emission using absorbed molecules via chemical reactions. Although these studies stand on their own merits, they have not considered the harvesting of molecules at the TX and recycling them via biochemical reactions for the following rounds of emission to reduce the energy cost.

\begin{figure*}[!t]
\begin{center}
\includegraphics[width=0.8\textwidth]{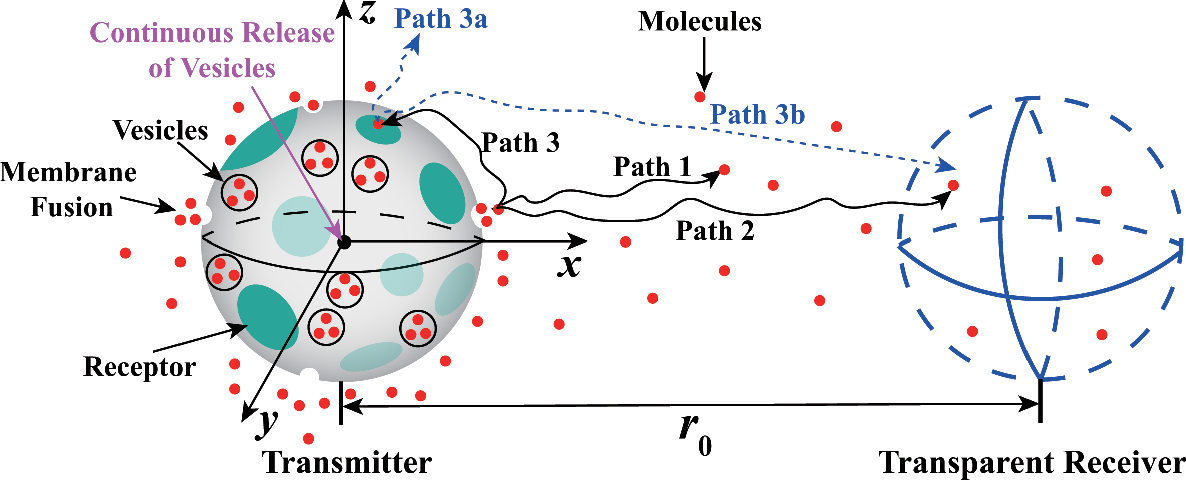}
\caption{Illustration of the MC system model where a spherical TX covered by heterogeneous receptors communicates with a transparent RX. Paths 1, 2, and 3 represent all possible diffusion paths of molecules after their release from the TX membrane.}\label{sys}
\end{center}
\end{figure*}
	
	Molecule harvesting is a known biological mechanism in synaptic communication. One typical example is reuptake \cite{rudnick1993synapse}, which is the re-absorption of neurotransmitters in the synaptic cleft by neurotransmitter transporters on the pre-synaptic neuron. This mechanism is necessary for synaptic communication since it allows to recycle neurotransmitters and control the time duration of signal transmission. Motivated by this biological example, a recent study \cite{ahmadzadeh2022molecule} developed a spherical molecule harvesting TX model where the membrane is equipped with receptors that can react with information molecules. Specifically, the authors considered a homogeneous TX surface, where they assumed an infinite number of receptors cover the entire TX surface or a finite number of receptors with identical sizes are uniformly distributed over the TX surface. However, the analysis in \cite{ahmadzadeh2022molecule} may become inaccurate in practice as the receptors on the TX surface may have different sizes and arbitrary locations. In particular, receptor clustering \cite{duke2009equilibrium} is one phenomenon that leads to heterogeneous receptors on the cell membrane. Thus, in this paper, we consider a spherical TX, whose membrane is covered by heterogeneous receptors that may have different sizes and arbitrary locations, and we assume each receptor to be fully absorbing.
Moreover, we model the transportation of molecules within the TX based on \cite{huang2021membrane}, where molecules are encapsulated within vesicles and are later released from the TX through the fusion of the vesicle and the TX membrane. Rather than assuming an impulsive release of vesicles at the TX center as in \cite{huang2021membrane}, we model a more realistic scenario involving the continuous generation of vesicles within the TX. Furthermore, we consider a transparent receiver (RX) and investigate the channel impulse response (CIR), where we consider that molecules may degrade when they propagate from the TX to the RX. We emphasize that the analysis in \cite{ahmadzadeh2022molecule} is not applicable when considering heterogeneous boundary conditions and vesicle-based release at the TX.
	
	The main contributions of this paper can be summarized as follows. We derive the rate at which molecules are release from the TX membrane for the case of a continuous generation of vesicles within the TX. We also derive the fraction of molecules absorbed at the TX and the probability that a released molecule is observed at the RX, where all expressions are functions of the size and location of each receptor. Particle-based simulations (PBSs) are used to verify the accuracy of our expressions. Our numerical results reveal that the total number of molecules absorbed by the TX is not affected by the vesicle generation rate, and that evenly distributed receptors on the TX membrane can capture a greater number of molecules compared to randomly distributed receptors or a single receptor.
	
\section{System Model}

In this paper, we consider an unbounded three-dimensional (3D) environment where a spherical TX communicates with a transparent spherical RX, as depicted in Fig. \ref{sys}. We choose the center of the TX as the origin of the environment and denote radii of the TX and RX by $r_{\ss\T}$ and $r_{\ss\R}$, respectively.

\subsection{TX Model}

In this subsection, we present the TX model in terms of vesicle generation, molecule propagation, and the receptors on the TX membrane.
	\subsubsection{Vesicle Generation}
	For the considered TX model, we assume that type-$\sigma$ molecules are stored within
	and transported by vesicles, where each vesicle stores $\eta$ molecules. We assume that vesicles are continuously generated in the center of the TX. In particular, each vesicle is generated at a random time instant, and we approximate the generation process as a one-dimensional (1D) Poisson point process (PPP) \cite{barbour1988stein}. 
	We note that PPP has been frequently used to describe random biological processes \cite{keeler2016notes}. In particular, several previous studies, e.g., \cite{etemadi2019compound}, have modeled the generation of molecules as 1D PPP. Furthermore, we denote $N_\mathrm{v}$ as the total number of vesicles that the TX generates for a single transmission and $\mu\;[\mathrm{vesicles}/\mathrm{s}]$ as the average number of vesicles generated per second.
	\subsubsection{Transportation of Molecules}
	We assume that the TX is filled with a fluid medium that has uniform temperature and viscosity. After vesicles are generated in the center of the TX, they diffuse randomly with a constant diffusion coefficient $D_\mathrm{v}$ until they reach the TX membrane. Then, these vesicles fuse with the membrane to release the encapsulated molecules. According to \cite{huang2021membrane}, we model the membrane fusion (MF) process between the vesicle and the TX membrane as an irreversible reaction with forward reaction rate $k_\mathrm{f}\;[\mu\mathrm{m}/\mathrm{s}]$. Thereby, if a vesicle hits the TX membrane, it fuses with the membrane with probability $k_\mathrm{f}\sqrt{\pi\Delta t/D_\mathrm{v}}$ during time interval $\Delta t$ \cite{andrews2009accurate}. After MF, the molecules stored in the vesicles are instantaneously released into the propagation environment. In biological systems, cells maintain a balance in membrane length through the dynamic processes of exocytosis and endocytosis. Although these processes continually alter the membrane, the average cell membrane length remains unchanged over time. Therefore, in this paper, we assume a fixed size for the TX membrane.
	\subsubsection{Receptors on the TX Membrane}\label{r}
	We assume that there are $N_\mathrm{r}$ non-overlapping heterogeneous receptors distributed on the TX membrane, which may have different sizes and arbitrary locations. We assume the shape of the $i$th receptor to be a circle with radius $a_i$. We define $\mathcal{A}$ as the ratio of the total area of the receptors to the TX surface, i.e., $\mathcal{A}=\sum_{i=1}^{N_\mathrm{r}}a_i^2/(4r_{\ss\T}^2).$ In a spherical coordinate system, we denote $\vec{l}_i=[r_{\ss\T}, \theta_i, \varphi_i]$ as the location of the center of the $i$th receptor, where $\theta_i$ and $\varphi_i$ represent the azimuthal and polar angles of the $i$th receptor, respectively. As the receptors are non-overlapping, the locations and radii of the receptors satisfy $|\vec{l}_i-\vec{l}_j|\geq a_i+a_j, \forall i, j\in\{1,2,...,N_\mathrm{r}\}$.
	In this model, we assume that all receptors are fully absorbing and can only absorb the released type-$\sigma$ molecules. With this assumption, once a released diffusing type-$\sigma$ molecule hits a receptor, it is absorbed by the TX immediately. The absorbed molecules are then recycled through biochemical reactions for successive emission rounds, aiding in energy conservation. Furthermore, we assume that the TX membrane area that is not covered by receptors is perfectly reflective, which means that released diffusing type-$\sigma$ molecules are reflected back once they hit this area. In addition, as the protein catalyzing MF is different from the receptor responsible for molecule absorption, molecule release and absorption are treated as two independent processes.
	\subsection{Propagation Environment and RX Model}
	In this system, we consider a transparent spherical RX whose boundary does not impede the diffusion of molecules. The center of the RX is distance $r_0$ away from the center of the TX. We assume that the RX can perfectly count the number of molecules within its volume at time $t$ and use this value as the received signal. We also assume that the propagation environment between TX and RX is a fluid medium with uniform temperature and viscosity. Once molecules are released from the TX, they diffuse randomly with a constant diffusion coefficient $D_\sigma$. We further assume unimolecular degradation in the propagation environment, where type-$\sigma$ molecules can degrade to type-$\hat{\sigma}$ molecules that can neither be absorbed by the TX nor observed by the RX, i.e., $\sigma\stackrel{k_\mathrm{d}}{\longrightarrow}\hat{\sigma}$ \cite[Ch. 9]{chang2005physical}, where $k_\mathrm{d}\;[\mathrm{s}^{-1}]$ is the degradation rate constant.
	
	\section{Analysis of Release and Harvest of Molecules at TX}\label{aor}
	In this section, we first analyze the release of molecules from the TX when jointly considering the continuous generation of vesicles and the MF process at the TX membrane, and derive the molecule release rate from the TX membrane. We define the molecule release rate as the probability that molecules stored in vesicles, which were generated in the origin starting at time $t=0$, are released during the time interval $[t, t+\delta t]$ from the TX membrane. Here, $\delta t$ represents a very small value of time $t$. Second, we incorporate the effect of the heterogeneous receptors on the TX membrane and analyze the absorption of molecules by the TX. We further derive the fraction of molecules absorbed by the TX until time $t$.
	\subsection{Molecule Release Rate from TX Membrane}
	We define $\tau=N_\mathrm{v}/\mu$ as the time duration during which the vesicles are generated. In our previous study \cite{huang2021membrane}, we assumed that vesicles are instantaneously generated in the center of the TX and provided the corresponding molecule release rate $f_\mathrm{r}(t)$ in \cite[Eq. (5)]{huang2021membrane}, where we took the MF process into account. In the following theorem, based on $f_\mathrm{r}(t)$, we derive the molecule release rate, denoted by $f_\mathrm{c}(t)$, when vesicles are continuously generated in the center of the TX.
	\begin{theorem}\label{t1}
		The molecule release rate from the TX membrane at time $t$, when vesicles are continuously generated starting at time $t=0$, is given by
		\begin{align}\label{bk}
			f_\mathrm{c}(t)=\left\{\begin{array}{lr}
				f_{\mathrm{c}, 1}(t), ~~\mathrm{if}\;0<t\leq\tau,
				\\
				f_{\mathrm{c}, 2}(t), ~~\mathrm{if}\;t>\tau,
			\end{array}
			\right.
		\end{align}
		where
		\begin{align}\label{fc}
			f_{\mathrm{c}, 1}(t)=&\frac{4r_{\ss\T}^2 k_\mathrm{f}\mu}{N_\mathrm{v}D_\mathrm{v}}\sum_{n=1}^{\infty}\frac{\lambda_nj_0(\lambda_nr_{\ss\T})}{2\lambda_nr_{\ss\T}-\mathrm{sin}(2\lambda_nr_{\ss\T})}\notag\\&\times\left(1-\exp\left(-D_\mathrm{v}\lambda_n^2t\right)\right)
		\end{align}
		and
		\begin{align}\label{fc2}
			f_{\mathrm{c}, 2}(t)=&\frac{4r_{\ss\T}^2 k_\mathrm{f}\mu}{N_\mathrm{v}D_\mathrm{v}}\sum_{n=1}^{\infty}\frac{\lambda_nj_0(\lambda_nr_{\ss\T})}{2\lambda_nr_{\ss\T}-\mathrm{sin}(2\lambda_nr_{\ss\T})}\notag\\&\times\left[\exp\left(-D_\mathrm{v}\lambda_n^2(t-\tau)\right)-\exp\left(-D_\mathrm{v}\lambda_n^2t\right)\right].
		\end{align}
		In \eqref{fc} and \eqref{fc2}, $j_0(\cdot)$ is the zeroth order spherical Bessel function of the first kind \cite{olver1960bessel} and $\lambda_n$ is obtained by solving $D_\mathrm{v}\lambda_nj_0'\left(\lambda_nr_{\ss\T}\right)=k_\mathrm{f}j_0\left(\lambda_nr_{\ss\T}\right)$
		with $j_0'(z)=\frac{\mathrm{d} j_0(z)}{\mathrm{d} z}$ and $n=1,2,...$.
	\end{theorem}
	\begin{IEEEproof}
		Please see Appendix \ref{A1}.
	\end{IEEEproof}
	
	\subsection{Molecule Harvesting at TX}\label{m}
	As vesicles are generated at the center of the TX, molecules are uniformly released from the TX membrane, i.e., the probability of molecule release is identical for any point on the TX membrane. Therefore, we first need to derive the number of molecules absorbed by the TX at time $t$ when these molecules are uniformly and simultaneously released from the TX membrane at time $t=0$, which is denoted by $H(t)$. To this end, we consider a scenario where molecules were uniformly released at time $t=0$ from the surface of a virtual sphere centered at the TX's center and having radius $d$, where $d\geq r_{\ss\T}$. The fraction of molecules absorbed by the TX in this scenario was derived in \cite[Eq. (5)]{huang2022analysis}. Based on this result, we present $H(t)$ in the following lemma.
\begin{table*}[!t]
\newcommand{\tabincell}[2]{\begin{tabular}{@{}#1@{}}#2\end{tabular}}
\centering
\caption{Summary of expressions for Calculating $\frac{1}{G_{\T}}$ \cite{lindsay2017first}.}\label{tab2}
\begin{tabular}{|c|c|}
\hline
\textbf{Expression}&\textbf{Size and Distribution of receptors}\\\hline
\tabincell{c}{$\frac{1}{G_{\T}}=\frac{2}{N_\mathrm{r}\overline{m}\kappa r_{\ss\mathrm{T}}}\Bigg[1+\frac{\kappa}{2N_\mathrm{r}\overline{m}}
				\ln\left(\frac{\kappa}{2}\right)\sum_{i=1}^{N_\mathrm{r}}m_i^2
				+\frac{\kappa}{N_\mathrm{r}\overline{m}}\bigg(\sum_{i=1}^{N_\mathrm{r}}m_is_i$\\$+2\sum_{i=1}^{N_\mathrm{r}}
				\sum_{j=i+1}^{N_\mathrm{r}}m_im_j\mathcal{F}(\vec{l}_i',\vec{l}_j')\bigg)
				+\left(\kappa\ln\left(\frac{\kappa}{2}\right)\right)^2\frac{\vartheta}{4N_\mathrm{r}\overline{m}}
				+\mathcal{O}\left(\kappa^2\ln\left(\frac{\kappa}{2}\right)\right)\Bigg].$}& Any size, any distribution\\
			\hline
			$\frac{1}{G_\mathrm{T}}=\frac{\pi}{N_\mathrm{r}\kappa r_{\ss\mathrm{T}}}\Bigg[1+\frac{\kappa}{\pi}\bigg(\ln(2\kappa)
			-\frac{3}{2}+\frac{4}{N_\mathrm{r}}\sum_{i=1}^{N_\mathrm{r}}\sum_{j=i+1}^{N_\mathrm{r}}
			\mathcal{F}(\vec{l}_i',\vec{l}_j')\bigg)+\mathcal{O}\left(\kappa^2\ln\left(\frac{\kappa}{2}\right)\right)\Bigg]$.& Identical sizes, any distribution\\
			\hline
			$\frac{1}{G_{\T}}\approx\frac{1}{r_{\ss\T}}\left(1+\frac{\pi}{N_\mathrm{r}\kappa}+\frac{\frac{1}{2}\ln(\kappa\sqrt{N_\mathrm{r}})+\ln 2-\frac{3}{2}}{N_\mathrm{r}}-2N_\mathrm{r}^{-\frac{1}{2}}+N_\mathrm{r}^{-\frac{3}{2}}\right).$& Identical sizes, evenly distributed\\
			\hline
			$\frac{1}{G_\mathrm{T}}=\frac{\pi}{\kappa r_\mathrm{T}}\bigg[1+\frac{\kappa}{\pi}\left(\ln(2\kappa)-\frac{3}{2}\right)-\frac{\kappa^2}{\pi^2}\left(\frac{\pi^2+21}{36}\right)+\mathcal{O}(\kappa^3\ln\kappa)\bigg].$& Single receptor\\
			\hline
		\end{tabular}
	\end{table*}
\begin{lemma}
		The fraction of molecules absorbed by the TX at time $t$, when the molecules are uniformly and simultaneously released from the TX membrane at time $t=0$, is given by
	\begin{align}\label{Gt}
		H(t)=&\frac{ w\mathrm{erf}(\sqrt{k_\mathrm{d}t})}{\sqrt{k_\mathrm{d}D_\sigma}}-\frac{w\gamma}{\zeta}\Big(\exp(\zeta t)\mathrm{erfc}\left(\gamma\sqrt{D_\sigma t}\right)\notag\\&+\gamma\sqrt{\frac{D_\sigma}{k_\mathrm{d}}}\mathrm{erf}\left(\sqrt{k_\mathrm{d}t}\right)-1\Big),
	\end{align}
	where $w=D_\sigma G_{\ss\T}/(r_{\ss\T}(r_{\ss\T}-G_{\ss\T}))$, $\gamma=1/(r_{\ss\T}-G_{\ss\T})$,  $\zeta=\gamma^2D_\sigma-k_\mathrm{d}$, $\mathrm{erf}(\cdot)$ is the error function, and $\mathrm{erfc}(\cdot)$ is the complementary error function. According to the definition and explanation in \cite{berg1977physics}, $G_{\ss\T}$ can be treated as the ``capacitance" of the TX, which is determined by the locations and sizes of the receptors. We note that $G_{\ss\T}$ measures the ability of the TX to absorb molecules given the distribution of the receptors. The expressions for $G_{\ss\T}$ for different distributions and sizes of receptors are summarized in Table \ref{tab2}, where $\kappa=\frac{a_1}{r_{\ss\mathrm{T}}}$, $m_i=\frac{2a_i}{r_{\ss\mathrm{T}}\kappa\pi}$, $\overline{m}=\frac{1}{N_\mathrm{r}}\sum_{i=1}^{N_\mathrm{r}}m_i$, $s_i=\frac{m_i}{2}\left(\ln\left(\frac{4a_i}{r_{\ss\mathrm{T}}\kappa}\right)-\frac{3}{2}\right)$, $\vartheta=\frac{\left(\sum_{i=1}^{N_\mathrm{r}}m_i^2\right)^2}{N_\mathrm{r}\overline{m}}-\sum_{i=1}^{N_\mathrm{r}}m_i^3$, $\mathcal{F}(\vec{l}_i', \vec{l}_j')=\left[\frac{1}{|\vec{l}_i'-\vec{l}_j'|}+\frac{1}{2}\ln|\vec{l}_i'-\vec{l}_j'|-\frac{1}{2}\ln\left(2+|\vec{l}_i'-\vec{l}_j'|\right)\right]$ with $\vec{l}_i'=\vec{l}_i/r_{\ss\mathrm{T}}$, and $\mathcal{O}(\cdot)$ represents the infinitesimal of higher order and is omitted during calculation.
	\end{lemma}
	\begin{IEEEproof}
		We denote $H_\mathrm{u}(t)$ as the fraction of molecules absorbed by the TX at time $t$, when the molecules are uniformly released from a virtual spherical surface at time $t=0$, which is given by \cite[Eq. (5)]{huang2022analysis}. $H(t)$ can be obtained from $H_\mathrm{u}(t)$ by setting $d\rightarrow r_{\ss\T}$, i.e., $H(t)=\lim\limits_{d\rightarrow r_{\ss\T}}H_\mathrm{u}(t)$. By substituting \cite[Eq. (5)]{huang2022analysis} into this expression, we obtain \eqref{Gt}.
	\end{IEEEproof}
	
	We then denote $H_\mathrm{e}(t)$ as the fraction of molecules absorbed by the TX at time $t$ when vesicles are continuously generated in the center of the TX starting at time $t=0$. We present $H_\mathrm{e}(t)$ in the following theorem.
	\begin{theorem}\label{t2}
		The fraction of molecules absorbed at the TX by time $t$ when vesicles are continuously generated in the center of the TX starting at time $t=0$ is given by
		\begin{align}\label{ge}
			H_\mathrm{e}(t)=f_\mathrm{c}(t)*H(t),
		\end{align}
		where $*$ denotes convolution, and $f_\mathrm{c}(t)$ and $H(t)$ are given in \eqref{bk} and \eqref{Gt}, respectively.
	\end{theorem}
	\begin{IEEEproof}
		When molecules are released from the TX membrane at time $u$, $0\leq u\leq t$, the fraction of these molecules that can be absorbed by the TX is given by $f_\mathrm{c}(u)H(t-u)$. Therefore, $H_\mathrm{e}(t)$ can be expressed as $H_\mathrm{e}(t)=\int_{0}^{t}f_\mathrm{c}(u)H(t-u)\mathrm{d}u$, which can be further rewritten as the convolution in \eqref{ge}.
	\end{IEEEproof}
	
		%

\section{Analysis of Received Signal at RX}\label{aors}
In this section, we derive expressions for the received signal at the RX in two steps. In the first step, we assume an absence of receptors on the TX surface and derive the received signal at the RX. In the second step, we determine the number of molecules that no longer arrive at the RX because they were absorbed by the receptors on the TX.
Based on these derivations, we finally obtain the received signal at the RX.
\subsection{Problem Formulation}\label{pf}
We classify all possible diffusion paths of molecules after their release from the TX membrane into three categories, namely path 1, path 2, and path 3, as shown in Fig. \ref{sys}. Path 1 is the path where molecules diffuse in the propagation environment at time $t$, path 2 is the path where molecules move into the RX at time $t$, and path 3 is the path where molecules hit a receptor on the TX surface at time $t$. If receptors are assumed to not exist, we can further divide path 3 into path 3a and path 3b. Path 3a is the path where molecules diffuse in the propagation environment at time $t$ after having hit the TX at time $u$, where $u\leq t$, and path 3b is the path where molecules move into the RX at time $t$ after having hit the TX at time $u$. Then, the received signal at the RX at time $t$ includes molecules from path 2 and path 3b. The presence of receptors causes the received signal at the RX to decrease, and this decrease is caused by the molecules associated with path 3b, which will no longer arrive at the RX. In light of this, we denote $P_{\ss\T}(t)$, $P_\mathrm{r}(t)$, and $P(t)$ as the probabilities that a released molecule is observed at the RX at time $t$ when receptors do not exist, when molecules move along path 3b, and when receptors exist, respectively. Accordingly, $P(t)$ can be calculated as
\begin{align}\label{pt}
	P(t)=P_{\ss\T}(t)-P_\mathrm{r}(t).
\end{align}
\subsection{Derivation of Received Signal}
We first assume there are no receptors on the TX membrane. Similar to the procedure in Section \ref{m}, to derive $P_{\ss\T}(t)$, we first derive the received signal at the RX when molecules are uniformly and simultaneously released from the TX membrane at time $t=0$, denoted by $P_\mathrm{u}(t)$. When a molecule is released from an arbitrary point $\alpha$ on the membrane of a spherical TX, the probability that this molecule is observed at the RX, denoted by $P_\alpha(t)$, is given by \cite[Eq. (27)]{noel2013using}
\begin{align}\label{pit}
	P_\alpha(t)=&\frac{1}{2}\left[\mathrm{erf}\left(\frac{r_{\ss\R}-r_\alpha}{\sqrt{4D_\sigma t}}\right)+\mathrm{erf}\left(\frac{r_{\ss\R}+r_\alpha}{\sqrt{4D_\sigma t}}\right)\right]\exp(-k_\mathrm{d}t)\notag\\&+\frac{1}{r_\alpha}\sqrt{\frac{D_\sigma t}{\pi}}\left[\exp\left(-\frac{(r_{\ss\R}+r_\alpha)^2}{4D_\sigma t}-k_\mathrm{d}t\right)\right.\notag\\&\left.-\exp\left(-\frac{(r_{\ss\R}-r_\alpha)^2}{4D_\sigma t}-k_\mathrm{d}t\right)\right],
\end{align}
where $r_\alpha$ is the distance between point $\alpha$ and the center of the RX. By taking the surface integral of $P_\alpha(t)$ over the TX membrane, we derive and present $P_\mathrm{u}(t)$ in the following lemma.
\begin{lemma}\label{l2}
	If there are no receptors on the TX membrane, the probability that a molecule is observed at the RX at time $t$, when these molecules were uniformly and simultaneously released from the TX membrane at time $t=0$, is given by
	\begin{align}\label{pu}
		&P_\mathrm{u}(t)=\frac{1}{8r_0r_{\ss\T}}\left[\xi_1(r_0-r_{\ss\T}, t)+\xi_1(r_{\ss\T}-r_0, t)-\xi_1(r_0+r_{\ss\T}, t)\right.\notag\\&\left.-\xi_1(-r_0-r_{\ss\T}, t)\right]+\frac{D_\sigma t}{2r_{\ss\T}r_0}\left[\xi_2(r_{\ss\T}+r_0, t)+\xi_2(-r_{\ss\T}-r_0, t)\right.\notag\\&\left.-\xi_2(r_0-r_{\ss\T}, t)-\xi_2(r_{\ss\T}-r_0, t)\right],
	\end{align}
	where
	\begin{align}
		&\xi_1(z,t)=\exp\left(-\frac{(r_{\ss\R}-z)^2}{4D_\sigma t}-k_\mathrm{d}t\right)\left(r_{\ss\R}+z\right)\sqrt{\frac{4D_\sigma t}{\pi}}\notag\\&+\left(r_{\ss\R}^2+2D_\sigma t-z^2\right)\mathrm{erf}\left(\frac{r_{\ss\R}-z}{\sqrt{4D_\sigma t}}\right)\exp\left(-k_\mathrm{d}t\right),
	\end{align}	
	and
	\begin{align}
		\xi_2(z,t)=\mathrm{erf}((r_{\ss\R}+z)/(4D_\sigma t)^{1/2})\exp(-k_\mathrm{d}t).
	\end{align}
\end{lemma}
\begin{IEEEproof}
	Following \cite[Appendix B]{huang2021membrane}, we derive $P_\mathrm{u}(t)$ by computing the surface integral of $P_\alpha(t)$ over the TX membrane, which is given by
	\begin{align}\label{put}
		P_\mathrm{u}(t)=\frac{1}{2r_{\ss\T}}\int_{-r_{\ss\T}}^{r_{\ss\T}}P_\alpha(t)\big|_{r_\alpha=\sqrt{r_{\ss\T}^2+r_0^2-2r_0x}}\;\mathrm{d}x.
	\end{align}
	By substituting \eqref{pit} into \eqref{put}, we obtain \eqref{pu}.
\end{IEEEproof}

Based on $P_\mathrm{u}(t)$, we present $P_{\ss\T}(t)$ in the following lemma.

\begin{lemma}
	If there are no receptors on the TX membrane, the probability that a released molecule is observed at the RX at time $t$, when the vesicles are continuously generated in the center of the TX starting from time $t=0$, is given by
	\begin{align}\label{pt1}
		P_{\ss\T}(t)=f_\mathrm{c}(t)*P_\mathrm{u}(t),
	\end{align}
	where $f_\mathrm{c}(t)$	and $P_\mathrm{u}(t)$ are given in \eqref{bk} and \eqref{pu}, respectively.
	\begin{IEEEproof}
		The proof is similar to the proof of Theorem \ref{t2}, and thus omitted here.
	\end{IEEEproof}
\end{lemma}

Next, we derive the probability that a released molecule is observed at the RX associated with path 3b. As we assume that $\mathcal{A}$ and $a_i$ are extremely small compared to the distance between the TX and RX, each receptor can be regarded as a point TX which releases molecules with a release rate that equals the hitting rate of molecules on this receptor at time $t$. As molecules are uniformly released from the TX membrane, the probability of molecules hitting any point on the receptors is the same. We denote $h_{\mathrm{e},i}(t)$ as the hitting rate of molecules on the $i$th receptor at time $t$, and express $h_{\mathrm{e},i}(t)$ as $h_{\mathrm{e},i}(t)=\frac{\mathcal{A}_i}{\mathcal{A}}h_\mathrm{e}(t)$, where $\mathcal{A}_i=a_i^2/(4r_{\ss\T}^2)$ is the ratio of the area of the $i$th receptor to the TX surface and $h_\mathrm{e}(t)$ is the total hitting rate of molecules on the TX membrane. We recall that $\mathcal{A}$ is the ratio of the total area of receptors to the TX surface, as mentioned in Section \ref{r}. Then, we are ready to derive $P_\mathrm{r}(t)$ and $P(t)$. We present $P_\mathrm{r}(t)$ and $P(t)$ in the following theorem.
\begin{theorem}\label{t3}
	The probability that a released molecule is observed at the RX at time $t$, when vesicles are continuously generated in the center of the TX starting from time $t=0$, is given by
	\begin{align}\label{ptf}
		P(t)=f_\mathrm{c}(t)*P_\mathrm{u}(t)-P_\mathrm{r}(t),
	\end{align}
	where
	\begin{align}\label{prt}
		P_\mathrm{r}(t)=\frac{f_\mathrm{c,d}(t)\!*\!H(t)}{\mathcal{A}}\!*\!\!\sum_{i=1}^{N_\mathrm{r}}\!\mathcal{A}_iP_\alpha(t)\big|_{r_\alpha=d_i}.
	\end{align}
In \eqref{prt},	$f_\mathrm{c,d}(t)$ is given by
	\begin{align}\label{bkd}	
		f_\mathrm{c,d}(t)=\left\{\begin{array}{lr}
			\frac{4r_{\ss\T}^2 k_\mathrm{f}\mu}{N_\mathrm{v}}\sum_{n=1}^{\infty}\frac{\lambda_n^3j_0(\lambda_nr_{\ss\T})}{2\lambda_nr_{\ss\T}-\mathrm{sin}(2\lambda_nr_{\ss\T})}\exp\left(-D_\mathrm{v}\lambda_n^2t\right), \\~~~~~~~~~~~~~~~~~~~~~~~~~~~~~~~~~\mathrm{if}\;0<t\leq\tau,
			\\
			\frac{4r_{\ss\T}^2 k_\mathrm{f}\mu}{N_\mathrm{v}}\sum_{n=1}^{\infty}\frac{\lambda_n^3j_0(\lambda_nr_{\ss\T})}{2\lambda_nr_{\ss\T}-\mathrm{sin}(2\lambda_nr_{\ss\T})}\left[\exp\left(-D_\mathrm{v}\lambda_n^2t\right)\right.\\\left.-\exp\left(-D_\mathrm{v}\lambda_n^2(t-\tau)\right)\right], ~~\mathrm{if}\;t>\tau,
		\end{array}
		\right.
	\end{align}
 $d_i$ represents the distance between the center of the $i$th receptor and the center of the RX, i.e., $d_i=\sqrt{r_{\ss\T}^2-2r_0r_{\ss\T}\cos(\varphi_i)\sin(\theta_i)+r_0^2}$, and $f_\mathrm{c}(t)$, $P_\mathrm{u}(t)$, $H(t)$, and $P_\alpha(t)$ are given in \eqref{bk}, \eqref{pu}, \eqref{Gt}, and \eqref{pit}, respectively.
\end{theorem}
\begin{IEEEproof}
	Please see Appendix \ref{A4}.
\end{IEEEproof}

\section{Numerical Results}
\begin{figure}[!t]
	\begin{center}
		\includegraphics[width=0.9\columnwidth,height=2.35in]{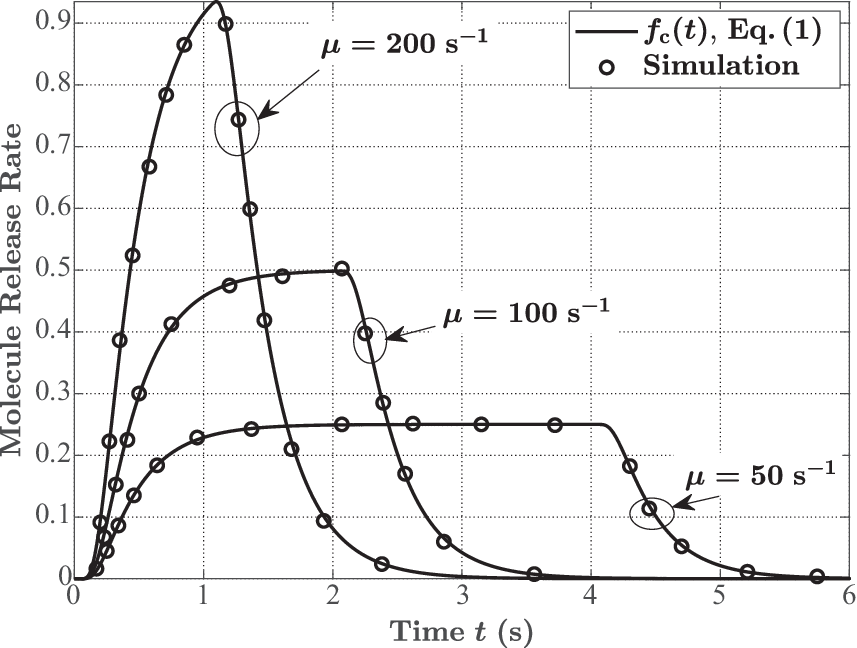}
		\caption{Molecule release rate from the TX versus time $t$ for different values of $\mu$.}\label{mrs}
	\end{center}
\end{figure}

\begin{figure}[!t]
	\begin{center}
		\includegraphics[width=0.9\columnwidth,height=2.35in]{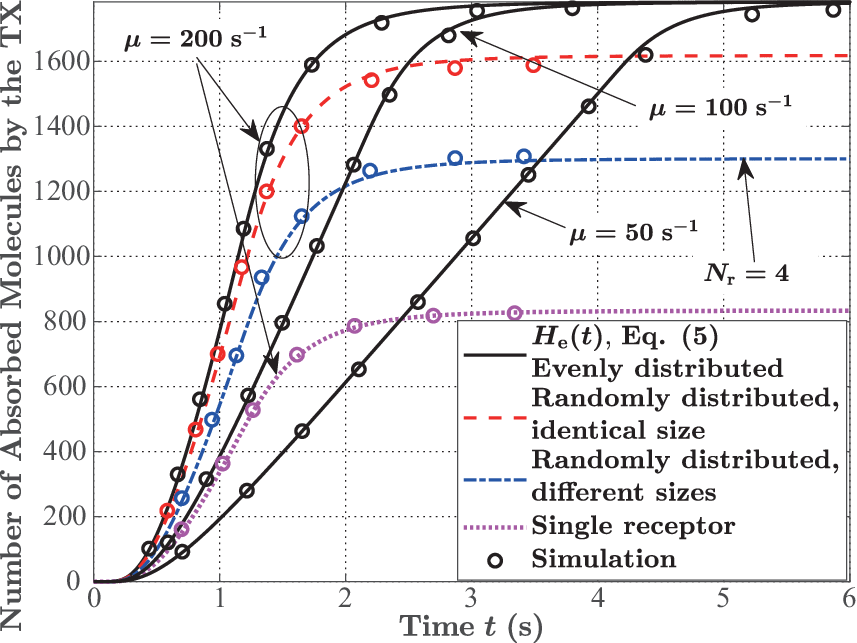}
		\caption{Number of molecules absorbed by the TX until time $t$ versus time $t$ for different distributions of receptors and different $\mu$, where $N_\mathrm{r}=11$.}\label{na4}
	\end{center}
\end{figure}

\begin{figure}[!t]
	\begin{center}
		\includegraphics[width=0.9\columnwidth,height=2.35in]{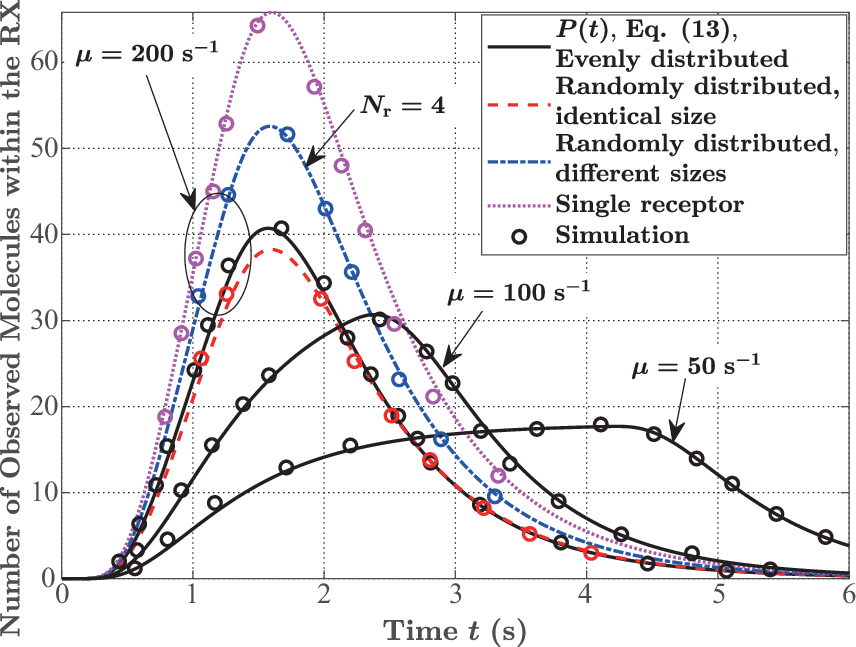}
		\caption{Number of observed molecules within the RX at time $t$ versus time $t$ for different distributions of receptors and $\mu$, where $N_\mathrm{r}=11$.}\label{v3}
	\end{center}
\end{figure}

In this section, we present numerical results to validate our theoretical analysis and offer insightful discussions. Specifically, we use PBSs to simulate the random diffusion of molecules. In our simulations, we model the vesicle generation process as a 1D PPP, resulting in the time interval between two consecutive generated vesicles following an exponential distribution with a mean of $1/\mu$. After vesicles are generated, they perform random diffusion with a variance of $2D_\mathrm{v}\Delta t_\mathrm{s}$ and fuse to the TX membrane with a probability of $k_\mathrm{f}\sqrt{\frac{\pi\Delta t_\mathrm{s}}{D_\mathrm{v}}}$, where $\Delta t_\mathrm{s}$ is the simulation step. The detailed simulation framework for modeling the diffusion of vesicles within the TX and the MF process at the TX membrane is illustrated in \cite[Sec. VI]{huang2021membrane}. After the molecules have been released from the TX, we record their positions in each simulation step. If the position of a molecule at the end of the current simulation step is inside the TX volume, we assume that this molecule has hit the TX membrane in this simulation step. The coordinates of the hitting points on the TX membrane are calculated by using \cite[Eqs. (36)-(38)]{huang2021membrane}. If the coordinates of the hitting point of a molecule are inside a receptor, we treat this molecule as an absorbed molecule. Otherwise, the molecule is reflected back to the position it was at the start of the current simulation step \cite{ahmadzadeh2016comprehensive}. We choose the simulation time step as $\Delta t_\mathrm{s}=10^{-6}\;\mathrm{s}$ and all results are averaged over 1000 realizations. Throughout this section, we set $r_{\ss\T}=5\;\mu\mathrm{m}$, $r_{\ss\R}=10\;\mu\mathrm{m}$, $N_\mathrm{v}=200$, $\eta=20$, $\mu=\{50, 100, 200\}\mathrm{s}^{-1}$, $D_\mathrm{v}=9\;\mu\mathrm{m}^2/\mathrm{s}$, $k_\mathrm{f}=30\;\mu\mathrm{m}/\mathrm{s}$, $\mathcal{A}=0.1$, $N_\mathrm{r}=\{1, 4, 11\}$, $r_0=20\;\mu\mathrm{m}$, $D_\sigma=79.4\;\mu\mathrm{m}^2/\mathrm{s}$, and $k_\mathrm{d}=0.8\;\mathrm{s}^{-1}$ \cite{hat2011b}, unless otherwise stated. From Figs. \ref{mrs}-\ref{v3}, we observe that the simulation results match well with the derived analytical curves, which validates our theoretical analysis in Sections \ref{aor} and \ref{aors}.

In Fig. \ref{mrs}, we plot the molecule release rate $f_\mathrm{c}(t)$ versus time $t$ for different values of $\mu$. We observe that when $\mu$ is small, $f_\mathrm{c}(t)$ maintains a constant value for a period of time. When $\mu$ is large, $f_\mathrm{c}(t)$ first increases and then decreases after reaching the maximum value. This is because small $\mu$ lead to long emission periods such that the concentration distribution of molecules within the TX becomes stable, which results in a stable molecule release rate.

\addtolength{\topmargin}{0.01in}

In Fig. \ref{na4}, we plot the number of molecules absorbed by the TX $N_\mathrm{v}\eta H_\mathrm{e}(t)$, versus time $t$, where we set $\mu\in\left\{50, 100, 200\right\}\mathrm{s}^{-1}$ and consider different numbers, distributions, and sizes of receptors. For receptors that are evenly distributed over the TX membrane, we apply the Fibonacci lattice \cite{gonzalez2010measurement} to determine the locations, which are given by \cite[Eq. (42)]{huang2022analysis}. For receptors that are randomly distributed over the TX surface, we consider receptors that either have the same size or different sizes. For receptors with different sizes, we set $N_\mathrm{r}=4$, their areas as $\mathcal{A}_1=0.01$, $\mathcal{A}_2=0.02$, $\mathcal{A}_3=0.03$, and $\mathcal{A}_4=0.04$, and their locations as $\vec{l}_1=[5\;\mu\mathrm{m}, \pi/2, \pi]$, $\vec{l}_2=[5\;\mu\mathrm{m}, \pi/2, \pi/2]$, $\vec{l}_3=[5\;\mu\mathrm{m}, \pi/2, 0]$, and $\vec{l}_4=[5\;\mu\mathrm{m}, \pi/2, 3\pi/2]$ as in \cite{huang2022analysis}. For the single receptor, we set the location as $\vec{l}_\mathrm{s}=[-r_{\ss\T}, 0, 0]$. First, when receptors are evenly distributed over the TX membrane, we observe that the total number of molecules that the TX can absorb is independent of the value of $\mu$, which means the fraction of molecules that can be recycled for subsequent transmissions does not depend on the generation rate of vesicles. Second, we observe that the number of molecules absorbed by evenly distributed receptors is larger than that absorbed by randomly distributed receptors or a single receptor. This occurs because evenly distributed receptors maintain an equal spacing, effectively covering the entire TX surface. Thus, the receptors have a higher probability of absorbing molecules.


In Fig. \ref{v3}, we plot the number of observed molecules within the RX $N_\mathrm{v}\eta P(t)$, versus time $t$, where the parameter setting for the locations and sizes of the receptors and the vesicle generation rate are the same as in Fig. \ref{na4}. First, when receptors have identical sizes and are evenly distributed on the TX membrane, we observe that a larger $\mu$ leads to a higher peak received signal. This is because a larger $\mu$ leads to a faster release of molecules into the propagation environment such that more molecules can be observed within the RX at the same time. Second, by considering both Fig. \ref{na4} and Fig. \ref{v3}, we observe that the received signal is weaker when the TX absorbs more molecules. This illustrates a trade-off between energy efficiency and error performance since a weaker received signal can result in a higher detection error at the RX.

\section{Conclusion}
In this paper, we investigated molecule harvesting for a spherical TX covered by heterogeneous receptors of different sizes and at arbitrary locations. By considering a continuous generation of vesicles within the TX and a transparent RX, we derived the molecule release rate and the fraction of molecules that are absorbed by the TX. We also derived the probability that a released molecule is observed at the RX. Simulations verified our analysis. Our numerical results showed that the vesicle generation rate determines the peak value of the received signal, and that evenly distributed receptors on the TX membrane can capture more molecules compared to randomly distributed receptors or a single receptor. Future directions of this research include determining the maximum fraction of molecules that can be absorbed by the TX while still achieving a desired error performance target.
\appendices
\section{Proof of Theorem \ref{t1}}\label{A1}
We first consider $0<t\leq\tau$. When a vesicle is generated in the center of the TX at time $u$, $0\leq u\leq\tau$, the probability that this vesicle fuses with the TX membrane at time $t$ is given by $f_\mathrm{r}(t-u)$. Here, the molecule release rate equals the vesicle fusion probability since MF guarantees the release of molecules into the propagation environment. Due to the continuous generation of vesicles, the number of vesicles fusing with the TX membrane during time interval $[t, t+\delta t]$ is given by $\mu\int_{0}^{t}f_\mathrm{r}(t-u)\;\mathrm{d}u$. As we define $f_\mathrm{c}(t)$ as the probability of a vesicle fusing with the TX membrane during an infinitesimally small time interval, we obtain $f_\mathrm{c}(t)$ as
\begin{align}\label{fcp}
	f_\mathrm{c}(t)=\frac{\mu}{N_\mathrm{v}}\int_{0}^{t}f_\mathrm{r}(t-u)\;\mathrm{d}u=\frac{\mu}{N_\mathrm{v}}\int_{0}^{t}f_\mathrm{r}(u)\;\mathrm{d}u.
\end{align}
By substituting \cite[Eq. (5)]{huang2021membrane} into \eqref{fcp}, we obtain \eqref{fc}.
Second, we consider $t>\tau$. For $t>\tau$, all considered vesicles have already been generated by the TX. Therefore, the number of vesicles fusing with the TX membrane during time interval $[t, t+\delta t]$ is given by $\mu\int_{0}^{\tau}f_\mathrm{r}(t-u)\;\mathrm{d}u$. Then, we obtain $f_\mathrm{c}(t)$ as
\begin{align}\label{fcp2}
	f_\mathrm{c}(t)=\frac{\mu}{N_\mathrm{v}}\int_{0}^{\tau}f_\mathrm{r}(t-u)\;\mathrm{d}u=\frac{\mu}{N_\mathrm{v}}\int_{t-\tau}^{t}f_\mathrm{r}(u)\;\mathrm{d}u.
\end{align}
By substituting \cite[Eq. (5)]{huang2021membrane} into \eqref{fcp2}, we obtain \eqref{fc2}.

\section{Proof of Theorem \ref{t3}}\label{A4}
We first derive $h_\mathrm{e}(t)$ by taking the derivative of $H_\mathrm{e}(t)$ with respect to $t$, given by $h_\mathrm{e}(t)=\frac{\partial H_\mathrm{e}(t)}{\partial t}=H(t)*\frac{\partial f_\mathrm{c}(t)}{\partial t}$. By substituting \eqref{bk} into $\frac{\partial f_\mathrm{c}(t)}{\partial t}$, we obtain \eqref{bkd}. We then derive $P_\mathrm{r}(t)$. We recall that $h_{\mathrm{e},i}(u)$ is the molecule release rate from the $i$th receptor at time $u$, and $P_\alpha(t-u)\big|_{r_\alpha=d_i}$ is the probability that a molecule is observed at the RX at time $t$ assuming this molecule was released from the $i$th receptor at time $u$. We then denote $P_{\mathrm{r},i}(t)$ as the probability that a molecule is observed at the RX at time $t$ assuming molecules were continuously released from the $i$th receptor and derive it as $P_{\mathrm{r},i}(t)=\int_{0}^{t}h_{\mathrm{e},i}(u)P_\alpha(t-u)\big|_{r_\alpha=d_i}\mathrm{d}u=h_{\mathrm{e},i}(t)*P_\alpha(t)\big|_{r_\alpha=d_i}$. Then, the probability that a molecule is observed at the RX assuming molecules were released from all receptors is given by $P_\mathrm{r}(t)=\sum_{i=1}^{N_\mathrm{r}}P_{\mathrm{r},i}(t)$. By substituting $P_{\mathrm{r},i}(t)$ into this expression, we obtain \eqref{prt}. Finally, by substituting \eqref{pt1} and \eqref{prt} into \eqref{pt}, we obtain \eqref{ptf}.
\section*{Acknowledgement}
The work of Yu Huang was supported by the National Natural Science Foundation of China under Grant 6220116.

The work of Robert Schober was (partly) funded by the German Research Foundation (DFG) under project SCHO 831/14-1.
\bibliographystyle{IEEEtran}
\bibliography{ref}
\end{document}